\documentclass[12pt,preprint]{aastex}
\shortauthors{Li J. Z.}
\shorttitle{Clustered star formation in the RMC}

\begin{document}

\title{Is the emerging massive cluster NGC 2244 a twin cluster?}
\author{Jin Zeng Li$^{1,2}$}
\affil{$^{1}$National Astronomical Observatories, Chinese Academy of 
           Sciences, Beijing 100012, China (E-mail: ljz@bao.ac.cn) \\
       $^{2}$Armagh Observatory, College Hill, Armagh BT61 9DG, N. Ireland, UK
}

\begin{abstract}

We present in this paper the first near infrared study of the young open cluster 
NGC 2244, which is well known for its partially embedded nature in the Rosette
Nebula. Based on the spatially complete 2 Micron All Sky Survey, the young
OB cluster indicates apparent substructures. It is surprisingly resolved into a 
compact core that matches well the congregation of massive OB stars in the optical, a satellite 
cluster at a distance of 6.6~pc in its west and probably a
major stellar aggregate resembling an arc in structure right below the core.
This infrared study provides various new updates on its nature 
of the young open cluster, including its central position, physical scale
and stellar population. 
A disk fraction of $\sim 20.5\pm2.8\%$ is achieved for its members with
masses above 0.8 $M_{\odot}$. NGC 2244 is hence a unique example for the study 
of embedded clusters.
\end{abstract}


\section{Introduction}

The partially immersed massive cluster NGC 2244 is one of the youngest open 
clusters in the Galaxy. It lies at a distance of $\sim$ 1.5 kpc and has an
evolutionary age of about 3 Myrs (Ogura \& Ishida 1981). Recent studies on
its members, the eclipsing binary V578 Mon (Hensberge, Pavlovski \& Verschueren 2000)
and the Ap star NGC 2244-334 with a strong magnetic field (Bagnulo, Hensberge \& 
Landstreet et al. 2004), suggest an age of $\sim$ 2 Myrs, consistent with that 
from photometric studies (Park \& Sung 2002).
NGC 2244 is famous for its excavation of the spectacular HII region, the Rosette 
Nebula. This young OB cluster is extensively studied both in the optical (Ogura \& Ishida 1981;
Park \& Sung 2002; Li, Wu \& Chen 2002; Li \& Rector 2004; Li 2003;) and in the
X-ray wavelength (Gregorio-Hetem et al. 1998; Berghofer \& Christian 2002; 
Townsley et al. 2003). Furthermore, an extensive burst of cluster formation 
was revealed by Li \& Smith (2005a) to be taking place in the south-east quadrant of the 
Rosette Molecular Complex (RMC). However, no near infrared study of 
the emerging young cluster NGC 2244 is as yet available, the incipience of which
could have imposed strong impacts on its ambient molecular cloud.

The advent of the spatially complete search of the near infrared sky by the Two
Micron All Sky Survey (2MASS) has undoubtedly brought us a revolutionary stage 
of the study of particularly protoclusters still embedded in their natal clouds.
This study of the young open cluster NGC 2244 is initiated for the following 
objectives:

(1) Arbitrary central positions of NGC 2244 were presented in the literature based 
on e.g. coordinates of one of its massive members (Ogura \& Ishida 1981) and
similarly thereafter.  An infrared study will be an effective way to offer 
a statistically accurate central position of the cluster, and in the meanwhile 
help obsolete the complicated and inaccurate designations related to NGC 2244 
and the Rosette Nebula in the literature.

(2) Due to its partially embedded nature, conjectures exist on its spatial 
appearance of the cluster as well as on the morphology and orientation of the Rosette Nebula.
Infrared studies can provide a more complete census of its stellar population and 
offer a direct view on the spatial distribution of the members of the cluster as
respect to the HII region it excavated. 
This could be an efficient way to make corresponding issues clear.

(3) Disk frequency is an important reference for the study of
disk evolution and of issues related to the formation of planetary systems in young 
open clusters. NGC 2244 is unique in that the vast majority of its medium to low-mass
members in their pre-main sequence of evolution are immersed in the fierce UV
radiation of the central young OB stars and, in the meanwhile, are suffering from 
strong stellar wind dissipation.
This study may contribute to our understandings of the emergence of currently 
embedded clusters, and to the knowledge of disk evolution.

(4) Among the numerous medium and low mass young stellar objects (YSOs) in the central 
excavated region
of the cluster, very few are found to be associated with surviving Herbig-Haro (HH) 
objects or jets (Li \& Rector, 2004; Li 2003). Unlike other known HH jets from 
YSOs, the Rosette HH1 \& HH2 jets are unexpectedly found to be in a high ionization
state. Nevertheless, disk materials of the Rosette HH1
source are evidenced to be in the process
of rapid dissipation by the intensive UV radiation from the massive OB stars
(Li et al. 2004). This study may further help to reveal the nature of the jet 
exciting sources and, possibly, to what extent these disks were stripped of material.


\section{Data Acquisition \& Analysis}

Both archived data of the 2MASS Point Source Catalogue (PSC) and the IRAS Sky 
Survey Atlases (ISSA) are retrieved via IRSA (http://irsa.ipac.caltech.edu/). 
For an introduction of the 2MASS and IRAS missions, please refer to 
corresponding explanatory supplements. The following sample selection criteria 
were employed to guarantee the reliability of the 2MASS data in use and of the 
results from follow-up analysis: (i) Each source extracted from the 2MASS 
PSC must have a certain detection in all three of the J, H and Ks bands. This 
will directly constrain the `rd-flg' (data reduction flag) to only 1, 2 or 3 
in each digit and the `ph-qual' (photometric quality) to only A, B, C or D for 
each band; (ii) We further restrict the Ks signal to noise ratio, K-snr, to 
above 15 to tightly constrain field stars in the control field to the main 
sequence and post-main sequence loci on the (J-H)--(H-Ks) diagram. Note that 
we need certain detection in at least one of the three bands in the source 
distribution presented by Figs. 1 \& 2, in order to offer as complete a census 
of the cluster members as possible. In the analysis of the Ks Luminosity
Function (KLF) of the cluster, we require instead only certain Ks detection.

\section{Spatial appearance of NGC 2244 in the near infrared}

Figure 1 shows the surface density map of the 2MASS sources toward NGC 2244 in a 
one square degree field centered at R.A.=06h31m55.9s, Dec=04d54m36s (J2000). 
The stellar density distribution is smoothed by a 2$\arcmin$\,$\times$\,2$\arcmin$ sampling box.
At an age of 2-3 Myrs (Park \& Sung 2002; Ogura \& Ishida 1981), NGC 2244 is strongly
evidenced to possess distinct substructures. It is resolved, in the near infrared,
into (i) an extended distribution of stellar sources associated with a compact core
in its south center of the excavated region, and (ii) an elongated substructure 
in the west indicating a clear 
density enhancement. This latter substructure also shows a distinct core at a 
distance of $\sim$ 6.6~pc to the west of the compact center of NGC 2244, possibly
indicating the existence of a prominent subcluster or rather a satellite cluster. 
Indeed, this is substantiated by the presence of an apparent congregation of 
X-ray emission sources in the ROSAT PSPC image of NGC 2244 (see Fig. 1 of Berghofer \& Christian
2002; Townsley et al. 2003).
The density map shows a physical scale of $\sim$ 2.2 pc ($\sim$5$\arcmin$) of the
condensed core of NGC 2244, consistent with the estimation based on the distribution 
of the central OB stars of the cluster (Townsley et al. 2003).

In Figure 2a, we construct the radial density profile of NGC 2244 centered at
R.A.=06h31m59.9s, Dec.=04d55m36s (J2000), the peak of the projected stellar density 
distribution.  We calculate the surface distribution of the sources using annuli 
having both constant areas and equal radial steps. The stellar distribution
is then compared to the averaged level of stellar density in the unreddened control 
fields (see Sect. 4).  An inverse radius model (n(r)=3.5+21.87/r, $\chi^2$=2.087) 
roughly describes the density distribution of the cluster.
The stellar density profile drops steeply and merges with the background density 
distribution at a radius of around 20$\arcmin$ or 8.1~pc at a distance of 1.5kpc,
which is commensurate with its extent both in the optical (Townsley et al. 2003) and
in the X-ray (Chen, Chiang \& Li 2004; Berghofer \& Christian 2002). Within this 
radius, the background distribution is subtracted and a net stellar population 
of 1901 $\pm$~75 of the cluster is achieved, taking into account statistical errors 
in the background stellar distribution as calculated based on the control fields.
Note that the control fields have the same area as the target field.
Note that the density profile also illustrates some oscillations likely due to the 
existence of substructures of the cluster.  
The radial density profile of the satellite cluster, centered at
R.A.=06h30m55.8s and Dec.=04d58m30s (J2000), is presented by Fig. 2b. It
does not match well an inverse radius fit due to probably its elongated distribution of
the cluster. A similar analysis gives an estimate of the radius of $\sim$3.1$\arcmin$ 
or 1.6~pc. The accumulative stellar population within this radius is 232 $\pm$~11.
The properties of the subclusters are summarized in Table 1. It should be noted 
that the stellar population for each subcluster given in Table 1 is derived based 
on calculations requiring only certain detection in one of the 3 bands and thus represents 
a complete census of the cluster members based on the 2MASS survey.

\begin{deluxetable}{lcccl}
\tabletypesize{\scriptsize}
\tablewidth{0pt}
\tablecaption{Properties of the clusters embedded in the Rosette Nebula}
\startdata
\noalign{\smallskip}
\tableline
\noalign{\smallskip}
Designation & Central Position & Physical Scale & Stellar Population & Peak density \\
            &  ---------------------  &     &      &  \\
            & R.A.(J2000) Dec.(J2000) & (pc) &  & arcmin$^{-2}$ \\
\noalign{\smallskip}
\tableline
\noalign{\smallskip}
NGC 2244 & 06 31 59.9~~~04 55 36 & $\sim$16~pc & 1901 $\pm$ 75 & 16.5 \\
Satellite cluster & 06 30 55.8~~~04 58 30 & $\sim$3.2~pc & 232 $\pm$ 11 & 13.75 \\
\noalign{\smallskip}
\enddata
\end{deluxetable}



The distribution of the known OB stars of NGC 2244 (Townsley et al.  2003) is 
presented in Fig. 3. It defines the center of the cluster in the optical but shows 
a more dispersed origin than the condensed core of the cluster in 
the near infrared that projected to the south center of the HII region (Fig. 1).
There has been an impression that the Rosette Nebula resembles a conical or cylindrical
hole penetrating the shells of the HII region and is oriented to the west of the 
line of sight (see the review by Townsley 
et al. 2003).  Its elongated appearance is, however, attributed by this study to 
the presence of the satellite cluster. 
The morphology of the excavated region of Rosette in the optical is consistent 
with the HII region having developed further to the west.
Based on the distribution of the massive OB stars and the appearance of the compact 
cluster in the infrared, the HII region appears rather to open at a direction some 
30$^{o}$ north of the line of sight. This results in an offset between the apparent 
center of the HII region (R.A.=06h31m55.9s, Dec.=04d59m56s J2000) and that of 
the embedded cluster.

\section{Color-Color Diagram}

  Color-color diagrams are taken as an effective way to judge the nature
of reddened stars and to determine individual extinction. We have constructed
(J-H)--(H-Ks) diagrams for two control fields (each has a radius of 20 arcmin,
commensurate with that of the target field).  One has a location to the south-west 
centered at (R.A.=06h30m00s, Dec.=03d45m00s, J2000) and the other to the 
north-east of the Rosette Nebula at (R.A.=06h35m30s, Dec.=05d47m00s, J2000), 
where visual extinction is expected to be quite low due to an investigation 
of the Digital Sky Survey (DSS) images. The diagram for the south-west field is presented 
in Fig. 4 as a comparison to that of the embedded cluster NGC 2244 and, on the other 
hand, for excluding the possible existence of foreground extinction.
It is clear that the 2MASS sources, selected based on the criteria introduced
in \S{2}, bound tightly to the main-sequence and post-main sequence loci and
show a typical mean H-Ks of about 0.2 mag. This indicates negligible
extinction toward the control field and presumably also in the foreground
of NGC 2244.

The (J-H)--(H-Ks) diagram of NGC 2244, within a radius of 20 arcmin centered
on the apparent center of the HII region,
is also presented in Fig. 4. A comparison of the source distribution in the 
diagram with that of the control field shows a distinct clump of background giants
situated in the left edge of the reddening band for normal stars, right above the 
post-main sequence track. 
This suggests a mean visual extinction of only about 1.5 mag toward the field of 
study, consistent well with an estimation of 1-2 mag in the literature (Celnik 1986).
All the 31 OB stars known to be associated with Rosette (Townsley et al. 2003)
are also put onto the diagram as a control sample, most of which have locations
at the bottom of the main sequence. Only a few of them indicate apparent foreground 
extinction with an upper limit of 3.5 mag, commensurate with the above estimation. 
Nevertheless, the low line of sight extinction is further supported by results 
from the CO \& C$^{13}$O maps (Blitz \& Stark 1986; Williams et al. 1995).
This indicates the HII region indeed has a cylindrical morphology and has probably
also penetrated the opposite side the Rosette Nebula and dissipated much
of the cloud materials in that direction.

Only one source shows an extinction of as large as 17 mag, it is located
in the satellite cluster.
A close investigation of the DSS and the 2MASS images of this source
indicates a highly obscured source probably located at the back
of a pillar object (Fig. 5).  With the exception of this source under a peculiar
environment, no source with a foreground visual extinction ${>}$ 10 mag was 
detected in the direction of NGC 2244.

Within our sample of 2760 sources, 206$\pm$20 fall off to the right of the reddening 
band for normal stars, and therefore show intrinsic color excess in the near infrared 
due to circumstellar dust emission. The uncertainty indicates the number of sources 
that fall within 1$\sigma$ in (H-Ks) from the excess line that defines the right edge 
of the reddening band for normal stars.
Some are situated further to the right of the locus of T Tauri stars, and are 
taken as strong candidates for Herbig Ae/Be stars. Indeed, three of them indicating
significant color excess correspond to known Herbig Ae/Be stars 
(Li, Wu \& Chen et al. 2002).  All pre-main sequence stars with a confirmed nature in
the literature are superimposed onto Fig. 4b and are marked with different symbols. 
Interestingly, 5 out of the 8 YSOs, including those 
associated with known optical jets (Li 2003; Li \& Rector 2004) and the two weak-line
T Tauri stars identified spectroscopically by Li et al. (2002), indicate locations 
associated with the unreddened main-sequence and post-main sequence tracks, and thus
show the lack of foreground extinction (or extinction from an envelope) and K-band 
excesses. This is just as expected for the weak line T-Tauri stars, which do not 
have disks (or at most may have remnant disks), but is very surprising for objects 
driving HH jets. Could this imply that the sources associated with the HH objects are 
not driving the flows? This explanation seems natural. However, (1) It is unlikely 
for any close companions to drive the HH flows because of the absence of apparent 
excess emission; (2) It should be noted that the jet driving sources in Rosette are 
immersed in strong UV ionization and dissipation from the nearby OB stars. We have 
accumulated evidence that these HH jets display many unique properties e.g. both the
Rosette HH1 and HH2 jets are found to have a high rather than a low ionization state
typical for other HH jets known in the literature including those identified in Orion 
(Reipurth et al. 1998). This might lead to different jet formation mechanisms, in which 
UV ionization could have played an important role in mass-loading onto the jet (Reipurth 
et al. 1998). Futhermore, model fitting to the spectral energy distribution of the HH1 
source resulted in a disk mass of only $\sim$0.006 M$\odot$ (Li \& Rector, 2004), smaller 
than the lower limit of 0.01 M$\odot$ around T Tauri stars. These interpret perhaps
their abnormal position of the jet driving sources on the (J-H)--(H-Ks) diagram and may 
suggest instead fast dissipation of the circumstellar disks of the YSOs, although color-color 
diagrams in the near infrared are deemed half complete in the census of circumstellar 
disks (Haisch, Lada \& Lada 2001). 



\section{Stellar population and disk frequency}

Due to its favorable orientation of the Rosette Nebula to the line of sight, 
and the apparently low interstellar extinction of the young open cluster that
immersed in the HII region, it is difficult to distinguish between unrelated 
field stars and especially the unreddened cluster members based on color
criteria. However, in order to offer a census of the stellar population of 
NGC 2244, the field contamination to the cluster members has to be estimated 
and corrected. Here we take the averaged stellar population of the control fields 
as a substitute to that of the field stars toward NGC 2244. Note that both the 
sample area of the cluster and the control fields have the same radius of 20 $\arcmin$. 
After subtraction of the field contamination of 1755 stars, we obtained a near 
infrared population of 1005 $\pm$ 42 of the cluster above a mass limit of $\sim$ 0.8 M$\odot$.


We took the sources with excess emission on the (J-H)--(H-Ks) diagram as
a lower limit census of the members of the cluster with circumstellar disks.
A disk fraction of $\sim 20.5\pm2.8\%$ is calculated for NGC 2244, taking into
account statistical uncertainties induced in the background subtraction. This 
is again restricted to the mass limit mentioned above. 

\section{Distribution of the excess emission sources}

Out of the 6576 sources that match our selection criteria within the one
square degree field toward NGC 2244, 450 are found to indicate excessive
emission in the near infrared, which is suggestive of the existence of
circumstellar disks. The surface density distribution of these
excess emission sources is smoothed by a 2$\arcmin$\,$\times$\,2$\arcmin$ 
sampling box and is presented by Fig. 6a.  Star formation in both a 
distributed and a clustered manner is evidenced toward this region. 
The compact core of NGC 2244 does 
correspond to an extended distribution of excessive emission sources,
however, it does not indicate distinct concentration toward its center as does
the satellite cluster. This is likely indicative of efficient disk dissipation 
in the center of NGC 2244.

Surprisingly, right to the south of the core of NGC 2244 lies what may be a separate
lane of sources unlikely due to simple projection effects. They form an arc in
structure in Fig. 6a, indicating probably their 
origin in inner or former swept-up layers of the HII region. Indeed, this 
structure is situated in the south boundary of the inner hole of the Rosette nebula
as presented by Fig. 6b, in which all excess sources are overplotted onto
the DSS R band image of Rosette. Each source is denoted as both a plus and a 
solid square. The size of the squares indicate the degree of Ks band excess. 
The arc structure is outlined by an enclosed continuous line in Fig. 6b, consistent 
with its distinct presence in the spatial distribution map of the 2MASS sources
constrained to (H-Ks) ${>}$ 0.5 mag (see Fig. 5 of Li \& Smith 2005a), which suggests
the existence of a separate congregate of YSOs or their association with 
differential extinction along the line of sight.
Nevertheless, the spatial appearance of the arc is roughly parallel to
the interaction layer of the Rosette Nebula and the ambient molecular cloud. 
This suggests a similar origin and fate of both the shell structures. The outer shell
is outlined by Regions A \& B (see Fig. 3) and 
other clumpy areas associated with recent star formation (Li \& Smith 2005b).
The apparent density enhancement in the south-east corner of Fig. 6b 
demarcates the presence of a proto-cluster in Region B of the outer shell.

\section{Color-Magnitude Diagram}

The sample sources of NGC 2244 that match the criteria introduced in \S{2}
are put onto the Color-Magnitude Diagram (CMD) illustrated in Fig. 7. The
31 OB stars known to be associated with Rosette (Townsley et al. 2003) are
marked with triangles, the majority of which meet well the unreddened
main sequence presented by a vertical line (Lejeune \& Schaerer 2001). 
As discussed in the above section, it is clear from the 
locations of the known pre-main sequence stars of NGC 2244 that foreground field stars
and cluster members with unnoticeable extinction are mixed together.
The gap usually exists between the unreddened main-sequence and the obscured
cluster members (Li \& Smith 2005b) is not evident in this specific case
due to the lack of interstellar extinction and/or the efficient gas dissipation of the cluster.


Based on the distribution of sources in the CMD, 17 sources that located
above the reddening vector of a A0 dwarf and with (H-Ks) ${>}$ 0.5 mag
are taken as candidate Herbig Be stars. Most of them exhibit excess
emission on the color-color diagram and are, therefore, good candidates 
for following up studies.

The strict sample selection 
method we employed along with the flux and resolution limited 2MASS 
survey, however, may result in the overlooking of sources well detected in H and 
Ks, but conceivably with the J band luminosity below the
detection limit due to differential extinction.
These sources are likely (1) shrouded by bright sources associated
with or happen to be projected nearby, or rather (2) severely suffering
from extensive emission in association or in the background. Fortunately, 
the interstellar extinction in NGC 2244 is substantially low and only a 
small number of sources are missed. They are tentatively included in Fig. 7 
and are presented with pluses. Some are likely YSOs situated in the 
surroundings of the compact cluster, where the extinction is predominately higher.

\section{Apparent Ks Luminosity Function}

KLF is widely used in studies of embedded clusters to provide clues on the 
investigation of initial mass function on a statistical basis.
In order to obtain the cluster KLF, field star contamination to the cluster
members has to be removed. The unreddened control field with the same area
is supposed to be representative of the background distribution of NGC 2244.
Extinction effects is not thought to affect severely our results due to
the low line of sight extinction toward NGC 2244.
All the sources with certain Ks detections within the sample area are used
to compute the raw KLF of the cluster (solid line in Fig. 8). KLF of the control 
field to the south-west of the Rosette Nebula is presented by the dot-dashed line 
as a comparison. The KLF of 
NGC 2244 is then constructed by subtracting that of the control field.
The KLF does not turn over till beyond the completeness limit of 2MASS 
(Ks=14.3 mag). It is found to match well a power law distribution
in the range between 7.0 ${<}$ Ks ${<}$ 14.0 mag. This gives a slope of 
$\sim$ 0.30 (Fig. 9), similar to 0.32 derived for NGC 2264 (Lada et al. 1993). 
However, it is somewhat lower than model predictions of around 0.4 for 
clusters with similar ages (Lada \& lada 1995).



\section{Summary and discussion}

The first study of the partially immersed young open cluster NGC 2244 in the near
infrared is carried out, based on the spatially complete 2MASS survey. This 
results in the following conclusions and implications.

(1) The young open cluster displays apparent substructures in the near infrared. It
is resolved into (i) a major cluster with a compact core, spatially consistent with 
that of the OB cluster extensively studied in the optical, and (ii) a distinct 
subcluster or satellite cluster located at $\sim$ 6.6~pc to the west of the condensed 
core. The south extension of the central cluster is further resolved into a separate 
lane of stars tracing an arc structure in the spatial distribution of the excessive 
emission sources toward this region. The appearance of the substructure discloses
the origin of the stars in 
an inner or former blown-up layer of the spectacular HII region. This congregation
of YSOs are too young to dissolve and merge with the major cluster. 

(2) This study of NGC 2244 provides various new updates on its nature of the cluster.
New centers, physical scales and stellar population of the subclusters revealed by 
this study are presented, which are thought to be statistically more accurate. Nevertheless, 
the long puzzling issue related to the morphology and orientation of the HII region 
of Rosette is made clear from an infrared view of the associated cluster(s). The
elongated appearance of the hollow region of Rosette is believed to be due to the
existence of the satellite cluster. The HII region is suggested to open instead 
at a direction some 30$^{o}$ north of the line of sight.

(3) Our study of the luminosity-limited sample further restrict the stellar population 
to a lower-limit of $\sim$ 1000, above a mass estimate of $\sim$ 0.8 M$\odot$.  A disk 
fraction of $\sim 20.5\pm2.8\%$ is then obtained based on the sources indicating intrinsic
excessive emission. This is consistent with the fraction of excess sources from JHK
studies of young clusters with similar ages (Haisch, Lada \& Lada 2001; Teixeira
et al. 2004; Baba et al. 2004), provided it is representative of that of the entire 
cluster.



{\flushleft \bf Acknowledgments~}

We are grateful to an anonymous referee for the constructive comments and suggestions
made, which improved a lot the scientific presentation of the paper.
This publication makes use of data products from the Two Micron All Sky Survey,
which is a joint project of the University of Massachusetts and the Infrared
Processing and Analysis Center/California Institute of Technology, funded by the
National Aeronautics and Space Administration and the National Science Foundation.
This work also makes use of the IRAS PSC \& ISSA data. Finally, we acknowledge
funding from SRF for ROCS, SEM.

\bibliographystyle{aa}

\clearpage

\figcaption[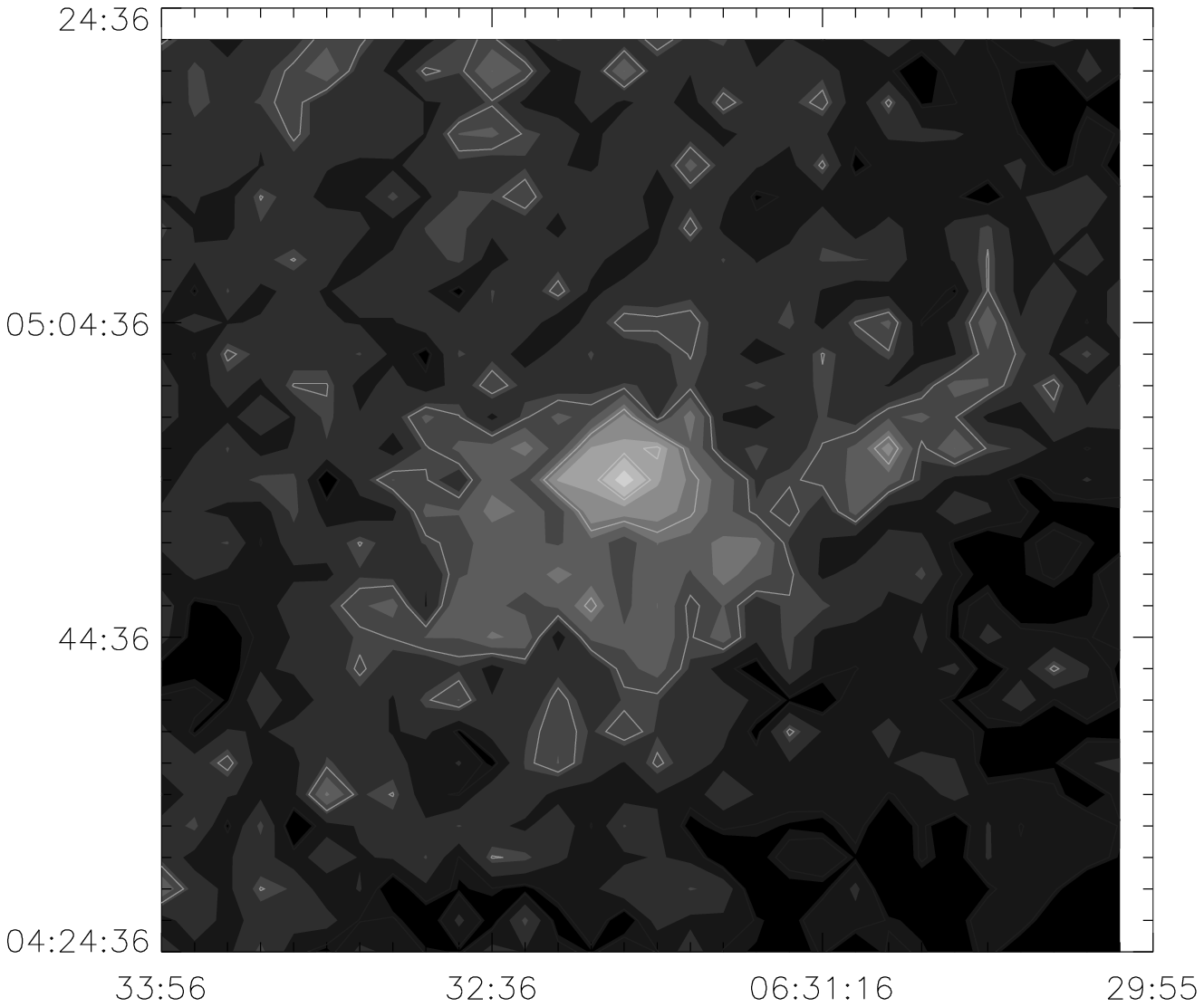]{The surface density distribution of the 2MASS sources with 
certain detection in at least one band towards NGC 2244 within one square degree. 
The stellar distribution is
filtered by a sample box with a width of 2$\arcmin$. Note about the apparent 
substructures indicated by the density map.}

\figcaption[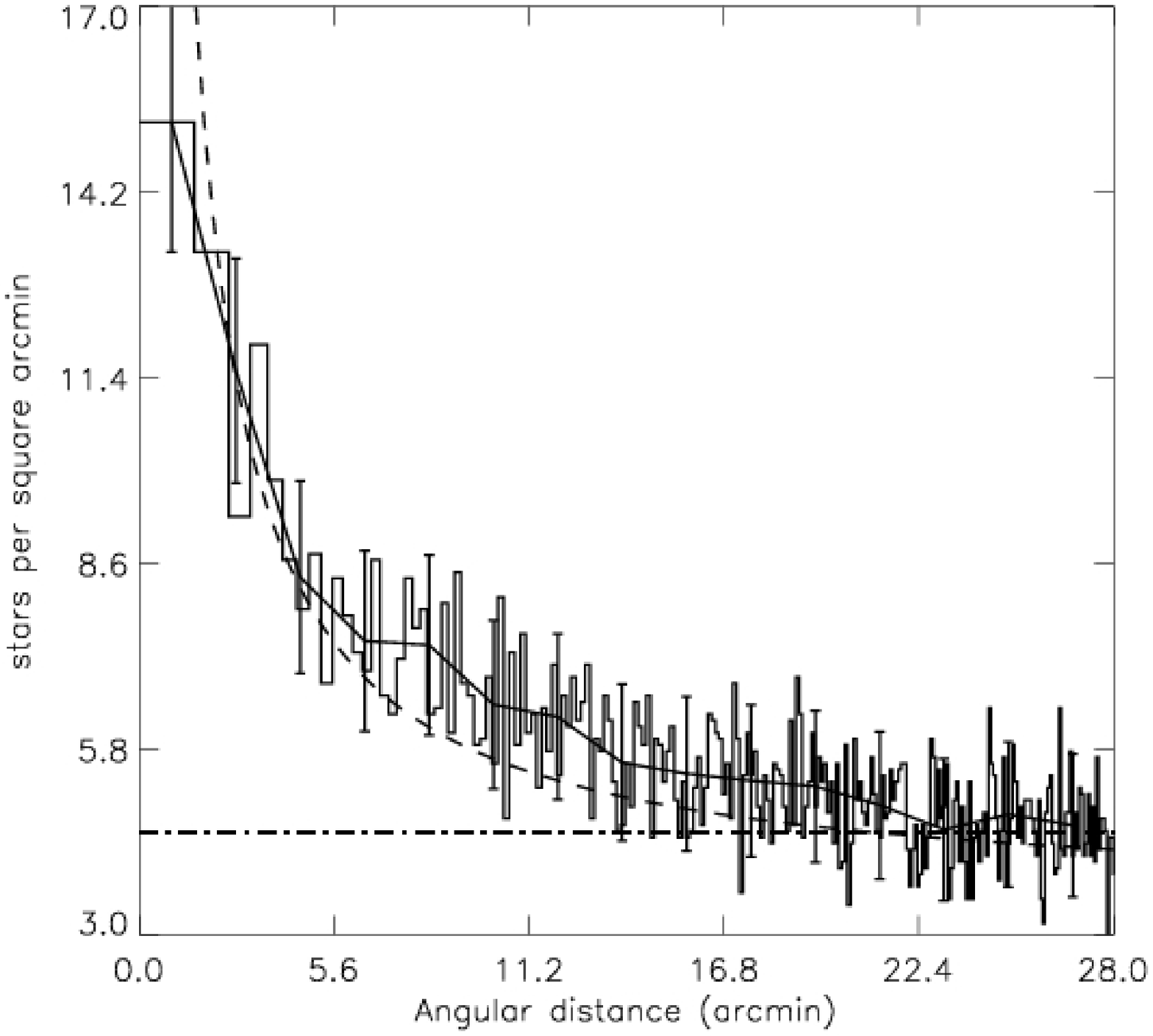,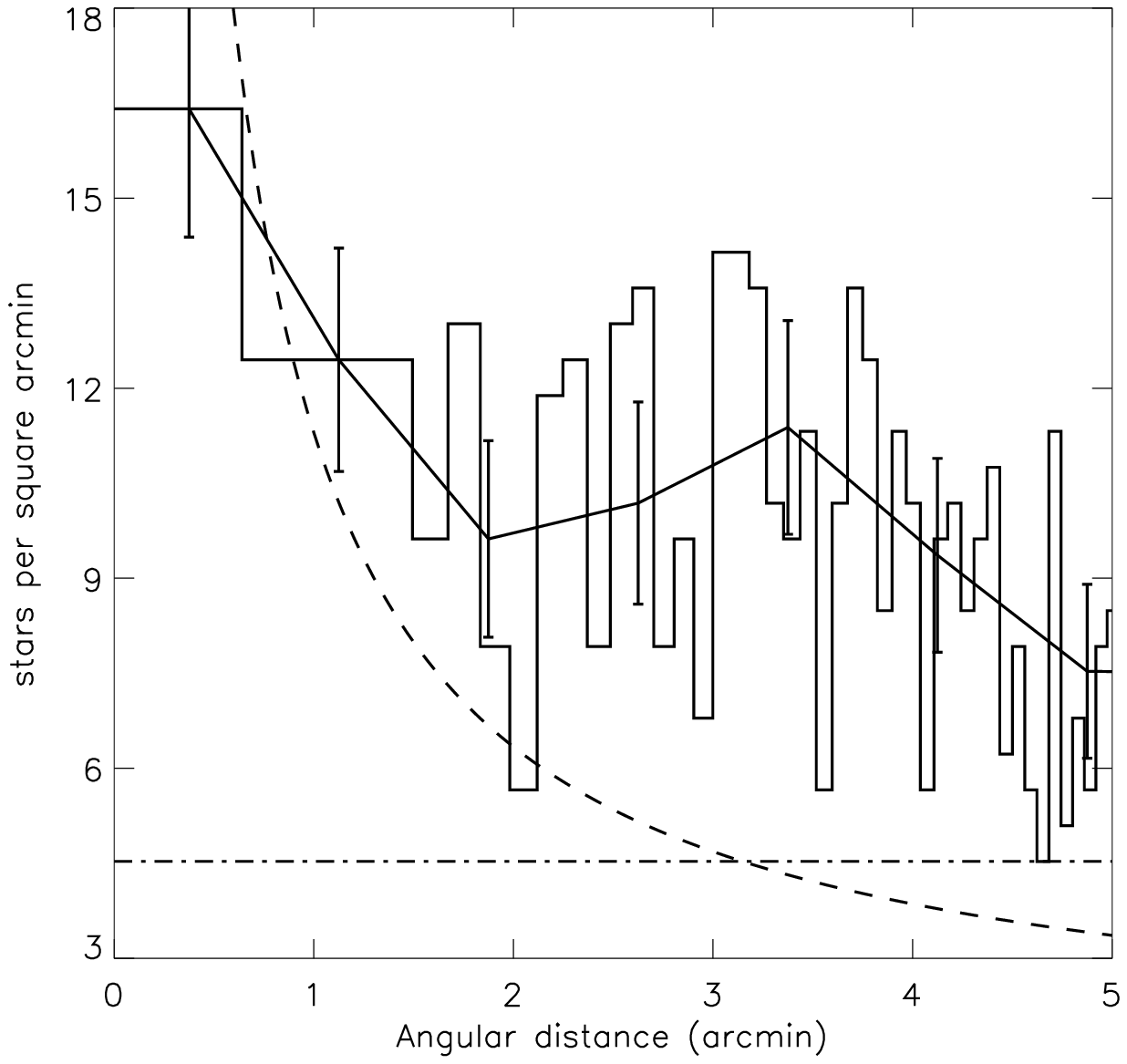]{Radial profiles of NGC 2244 and its satellite cluster.
a) The radial profile of NGC 2244 calculated using concentric annuli of both equal 
area and equal radial step. The width of the first annulus is R$_{0}$=1.85$\arcmin$; 
Error bars indicate statistical $\sqrt N$ error.  The dashed curve indicates a 
r$^{-1}$ fit to the corresponding radial stellar distribution.
b) The profile of the satellite cluster computed in the same way with 
R$_{0}$=0.75$\arcmin$.  As a comparison, the average surface density 
of the control fields is presented as a dot-dashed line.}

\figcaption[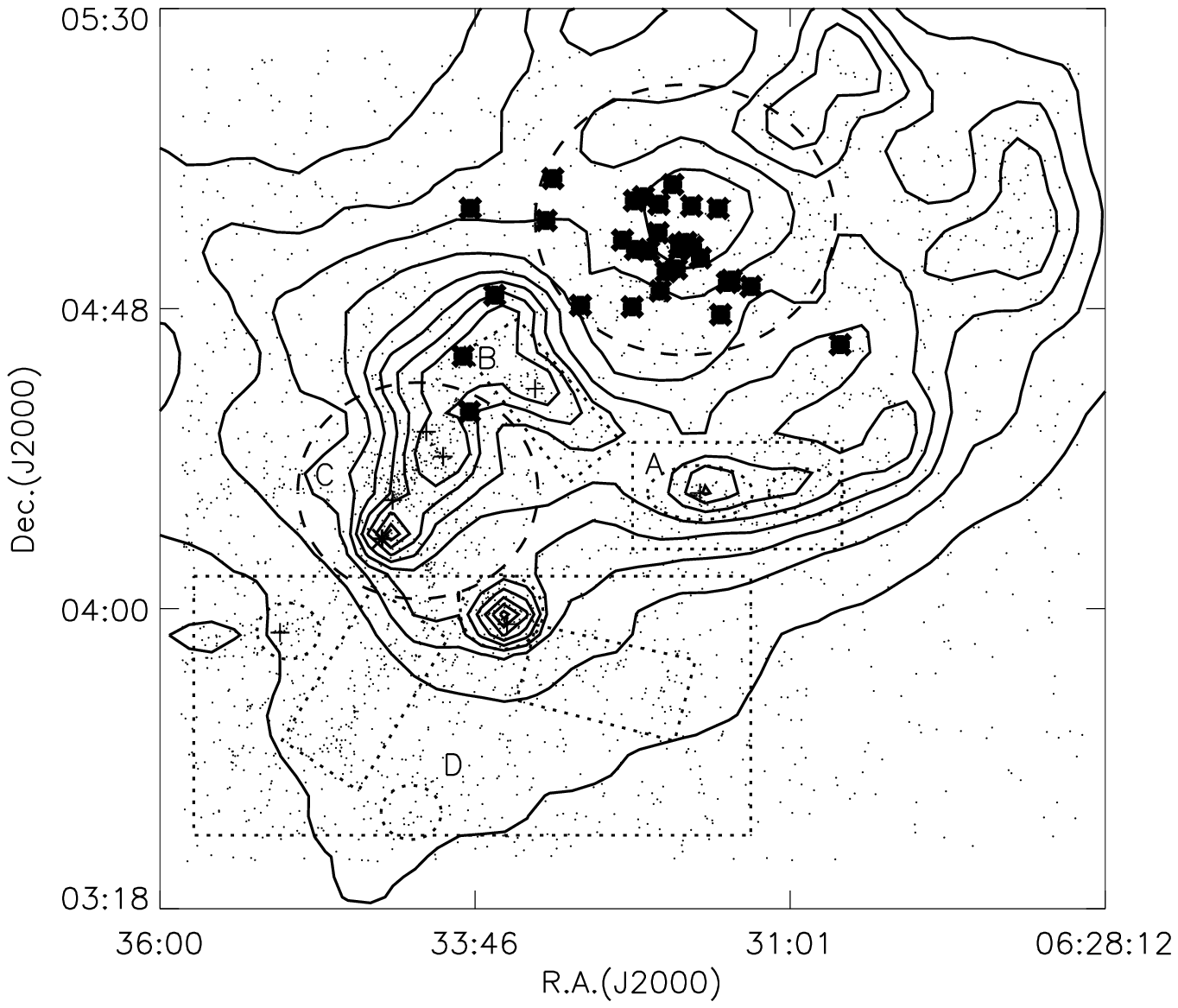]{Distribution of the OB cluster members of NGC 2244. Also overlapped is
the IRAS 100 $\mu$m emission of the RMC. Newly discovered near infrared clusters
to the south-east of NGC 2244 are outlined with dashed rectangles and circles (Li
\& Smith 2005a). All OB stars are indicated as asterisks in the plot.}


\figcaption[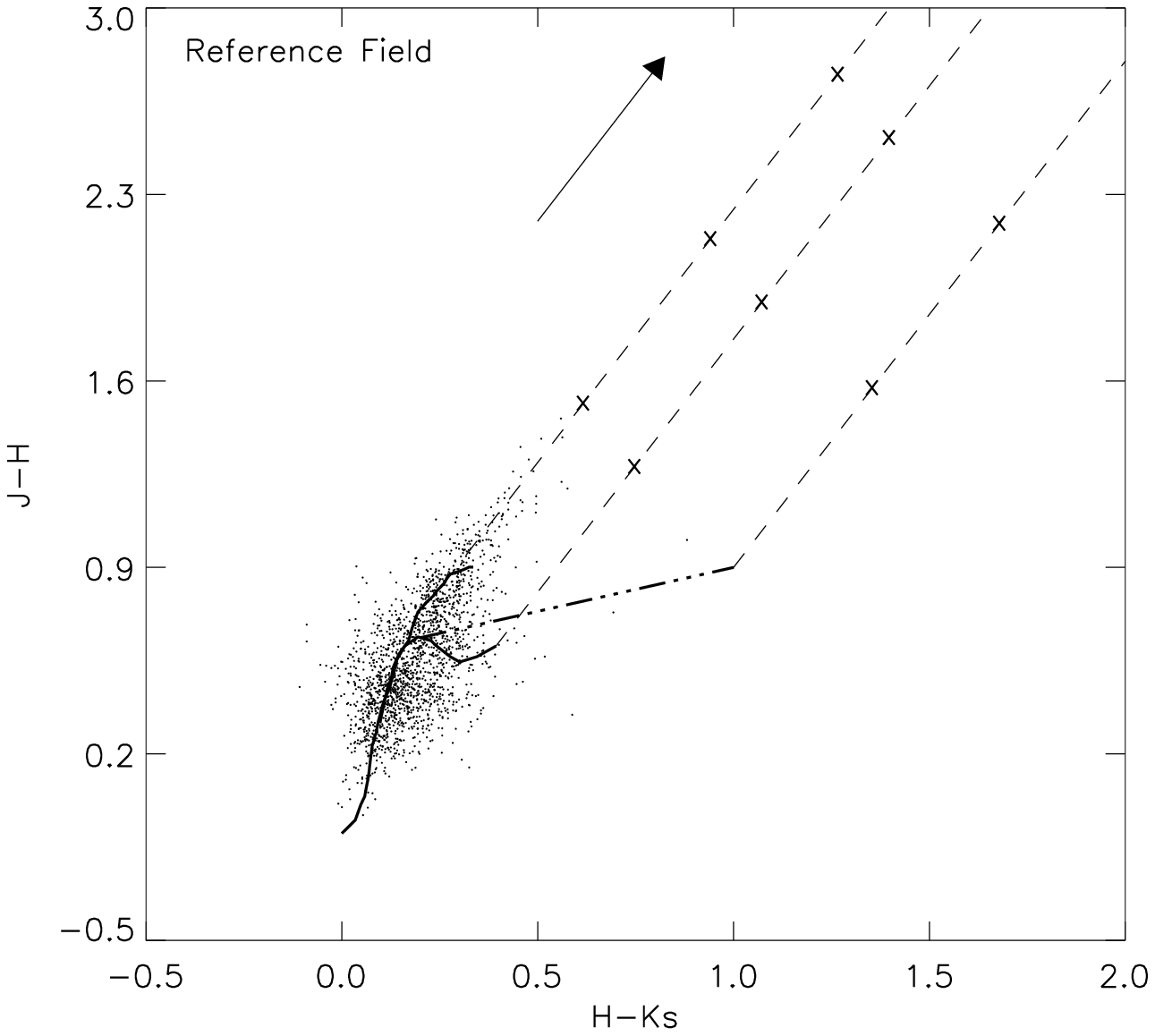,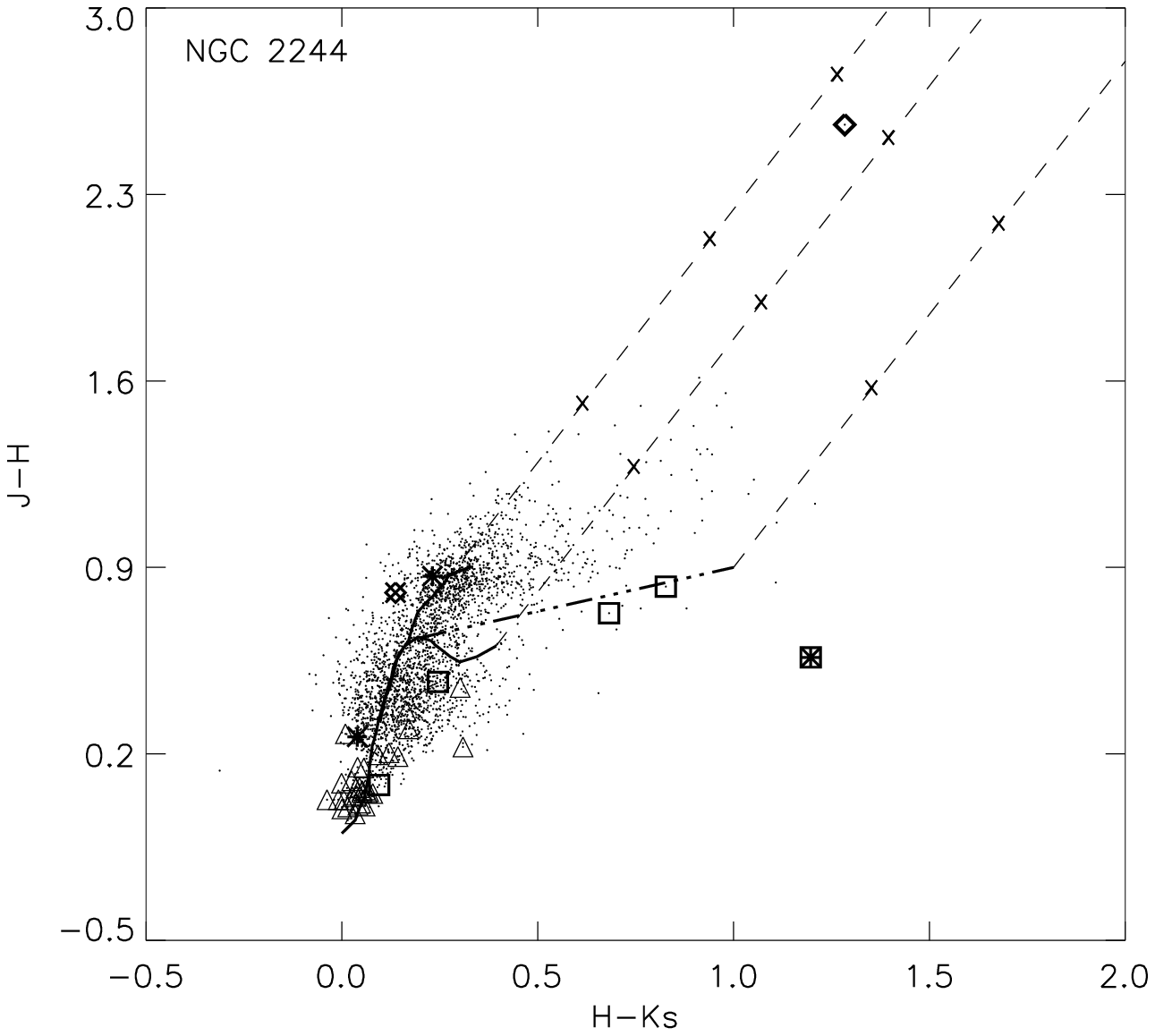]{The color-color diagram of the control field and that of 
NGC 2244, each with a radius of 20 arcmin.
The sample sources are plotted as dots. Pre-main sequence stars
with a confirmed nature are presented with different symbols. 
Asterisks indicate YSOs associated with optical jets and Herbig-Haro objects
(Li \& Rector 2004; Li 2003; Li et al. 2002). The companion
T Tauri star of the Rosette HH1 source is denoted as a cross in a diamond. Squares 
indicate Herbig Ae/Be stars and weak line T Tauri stars identified by spectroscopy
(Li, Wu \& Chen 2002). The 31 OB stars toward Rosette from Townsley et al. (2003) 
are shown as triangles.  Solid lines
represent the loci of the unreddened dwarves and giants (Bessel \& Brett 1988). The
arrow in the upper left of the plot shows a reddening vector of Av = 5 mag
(Rieke \& Lebofsky 1985). The dotted dashed line indicates the locus of
dereddened T Tauri stars (Meyer et al. 1997). The dashed lines define the
reddening band for normal stars and T Tauri stars, and are drawn parallel
to the reddening vector. Crosses are overplotted with an interval corresponding
to 5 mag of visual extinction.}

\figcaption[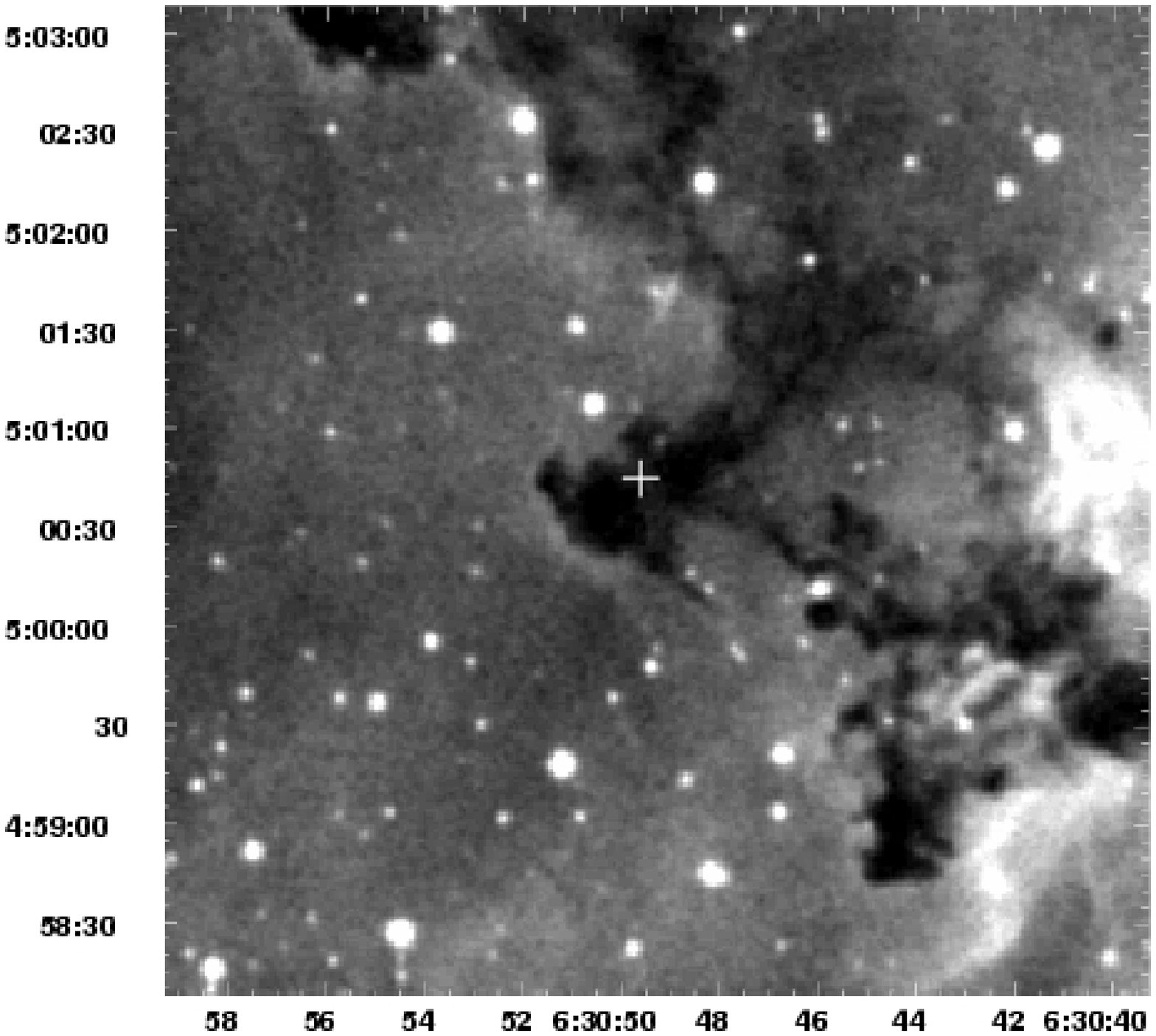,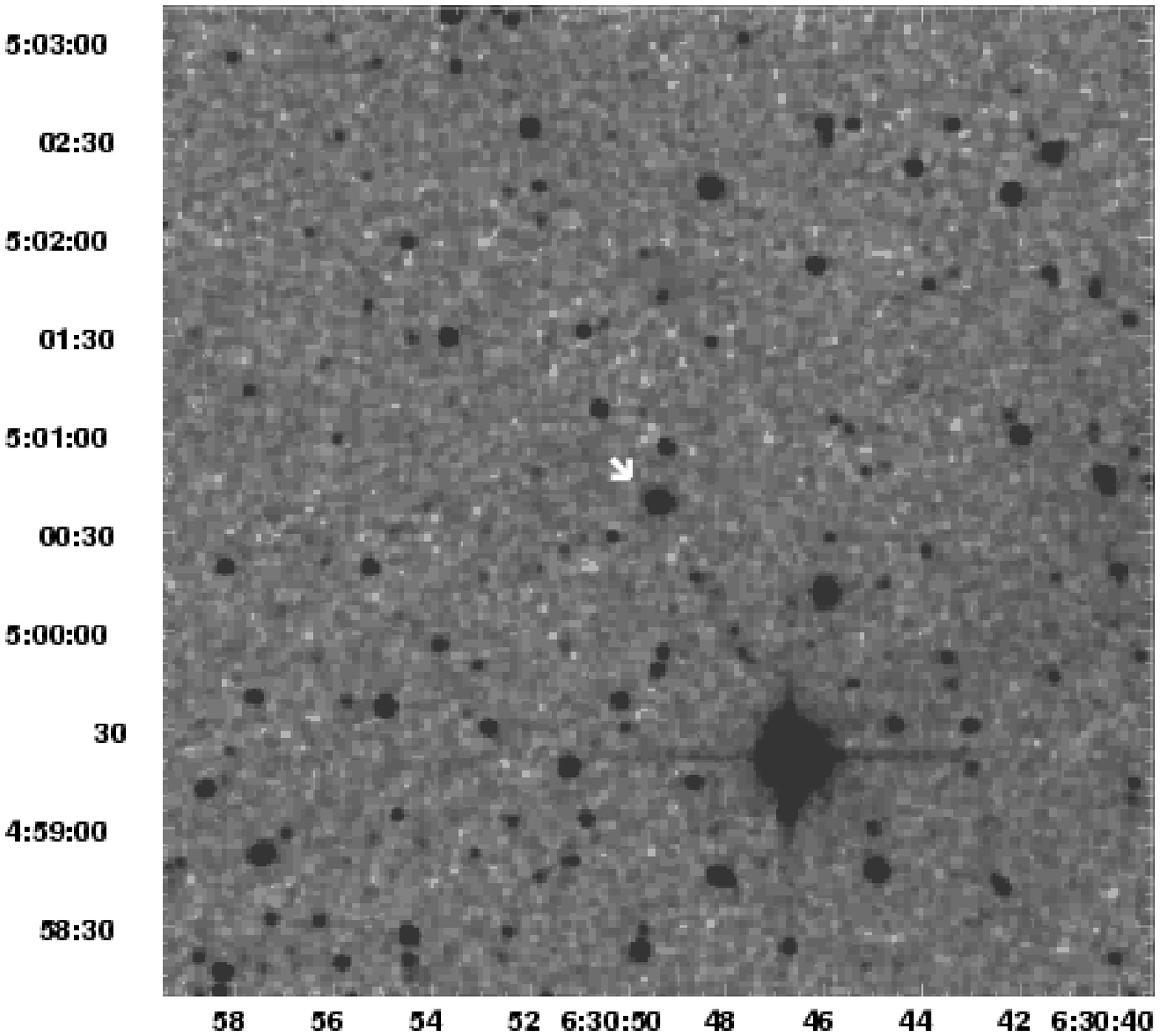]{DSS and 2MASS images of the source indicating the 
highest extinction, situated in the satellite cluster. It is heavily obscured
due to its location probably at the back of a pillar object.}

\figcaption[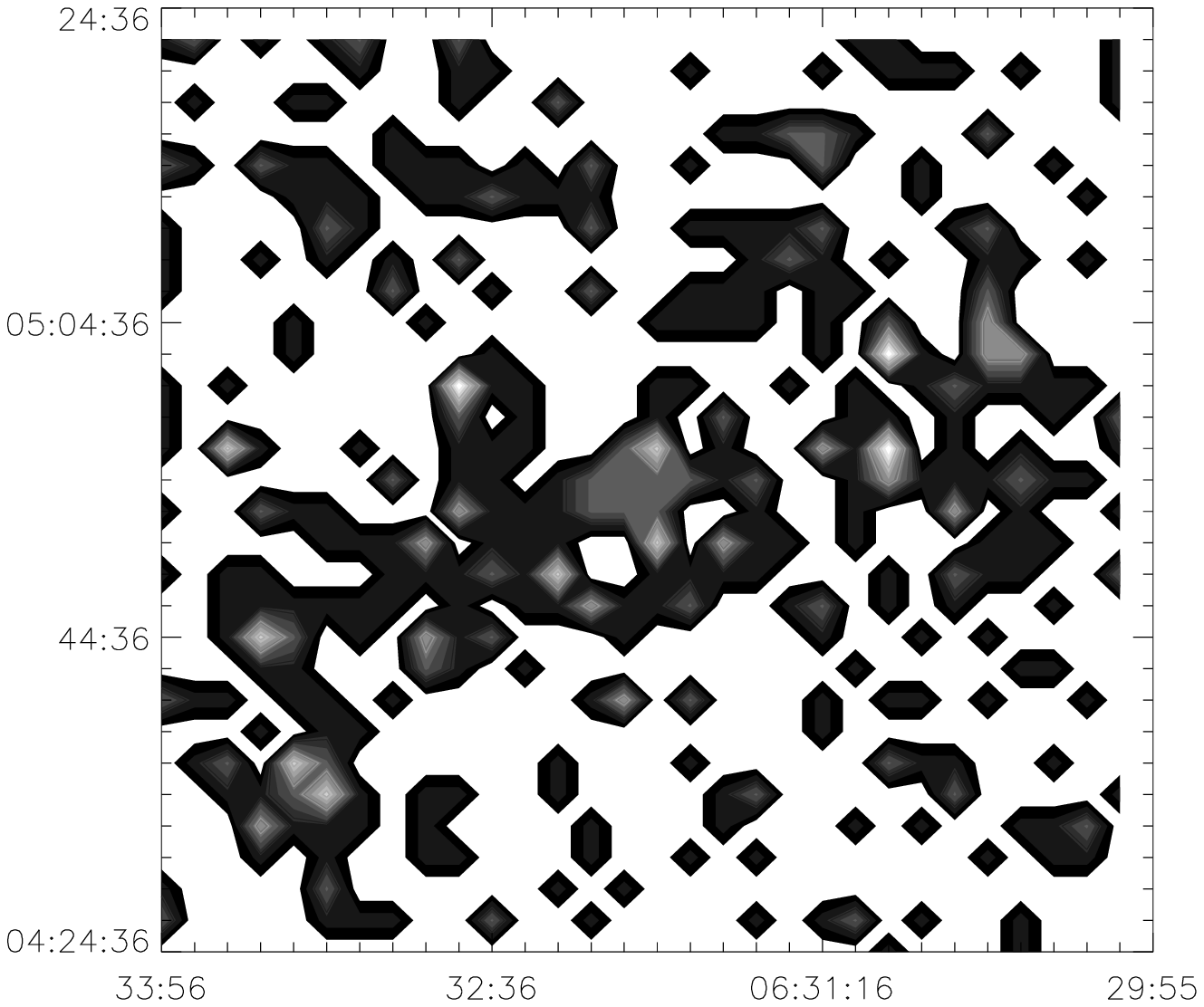,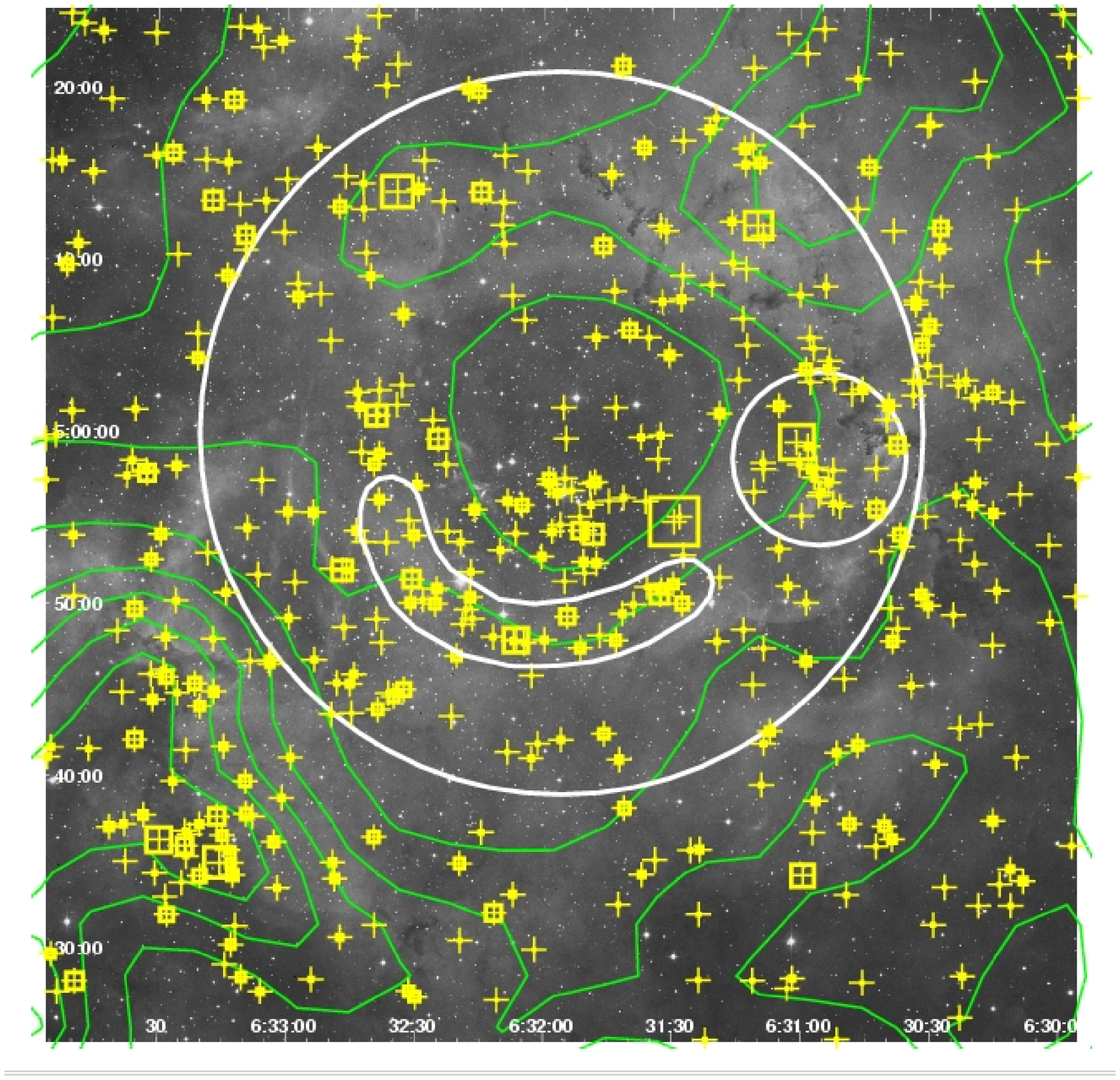]{Spatial distribution of the excessive emission sources
within the one square degree field toward NGC 2244. a) The surface density 
distribution of the excessive sources filtered by a sample box with a width of 
2$\arcmin$; b) The excessive sources overplotted onto the DSS R band image of
Rosette. Each source is plotted both as a plus and a solid square, where the
size of the square indicates the degree of Ks band excess. The large solid circle
with a radius of 20$\arcmin$ denotes the sample area of NGC 2244 and the smaller 
one outlines the position of the newly disentangled satellite cluster with a radius 
of 3.1$\arcmin$. Contours of the IRAS 100$\micron$ emission are also superimposed
onto the source distribution. The congregate of YSOs tracing an arc to the south
of the cluster core is enclosed by a continuous line. Note 
that the distinct density enhancements in the south-east corner corresponds
to Region B of the swept-up shell of the Rosette Nebula, which harbors also
recent star formation (Li \& Smith 2005b).}

\figcaption[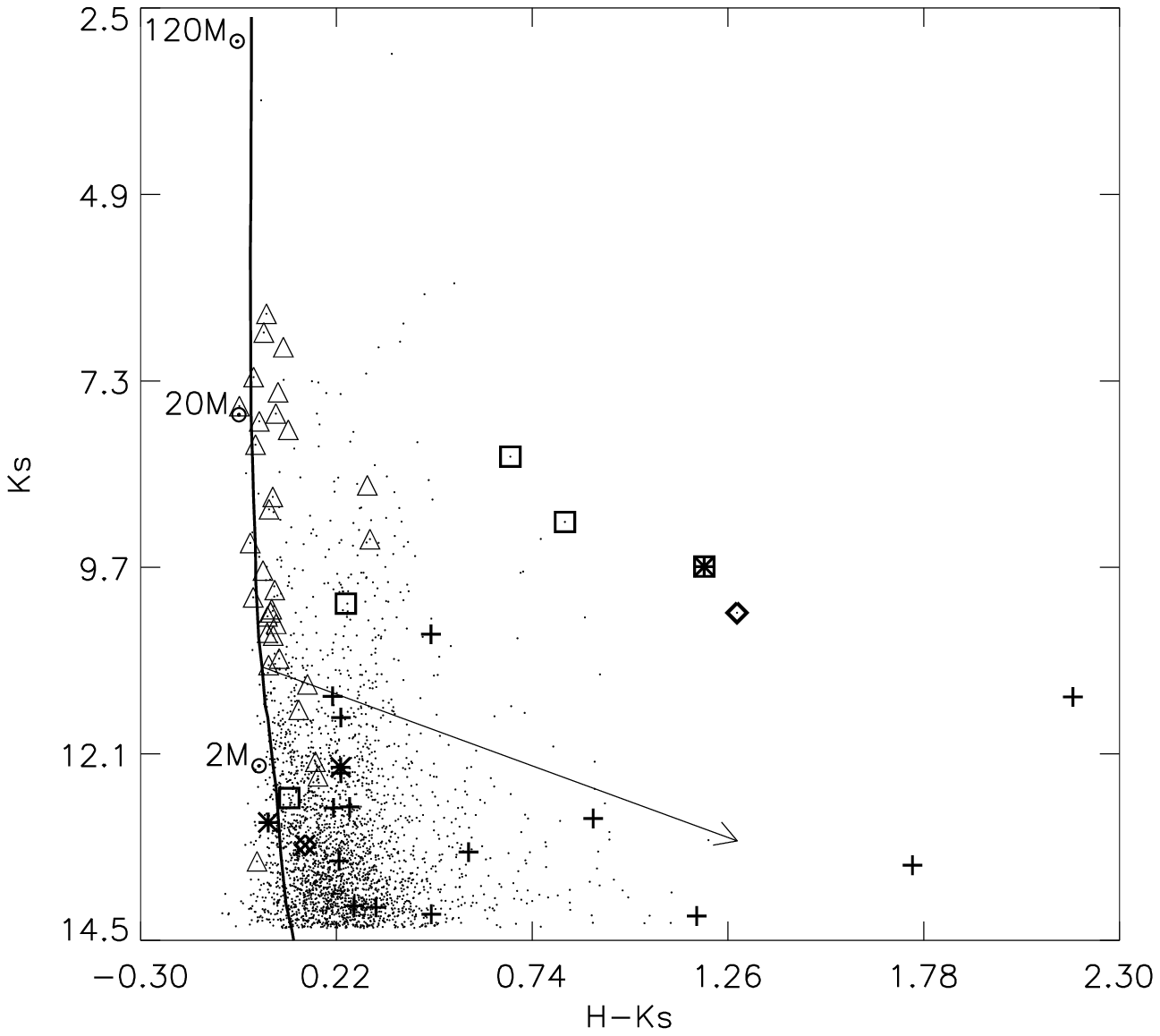]{The color magnitude diagram of NGC 2244. The vertical line
defines the unreddened main-sequence (Lejeune \& Schaerer 2001). The slanted line 
with an arrow at the tip denotes the reddening of Av = 20 mag of a A0 dwarf.
Pluses represent the criteria missed objects. Other symbols have the same 
indications as in Fig. 4.}

\figcaption[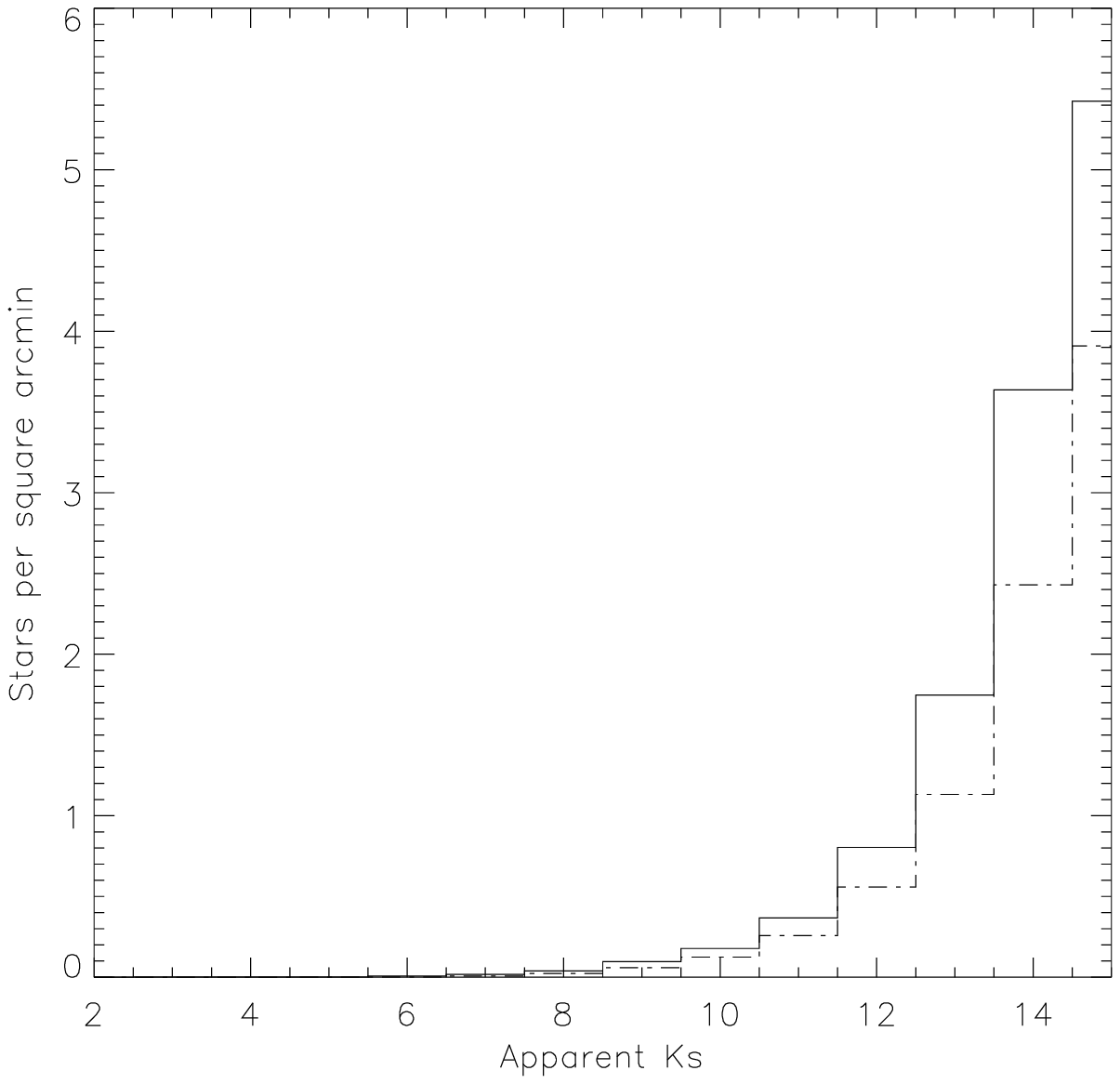]{The raw KLF of NGC 2244 as compared to that of the control 
field to the south-west of Rosette. The bin width is 1.0 mag. The solid histogram 
represents the raw KLF
of NGC 2244 and the dot-dashed line corresponds to that of the control field.}

\figcaption[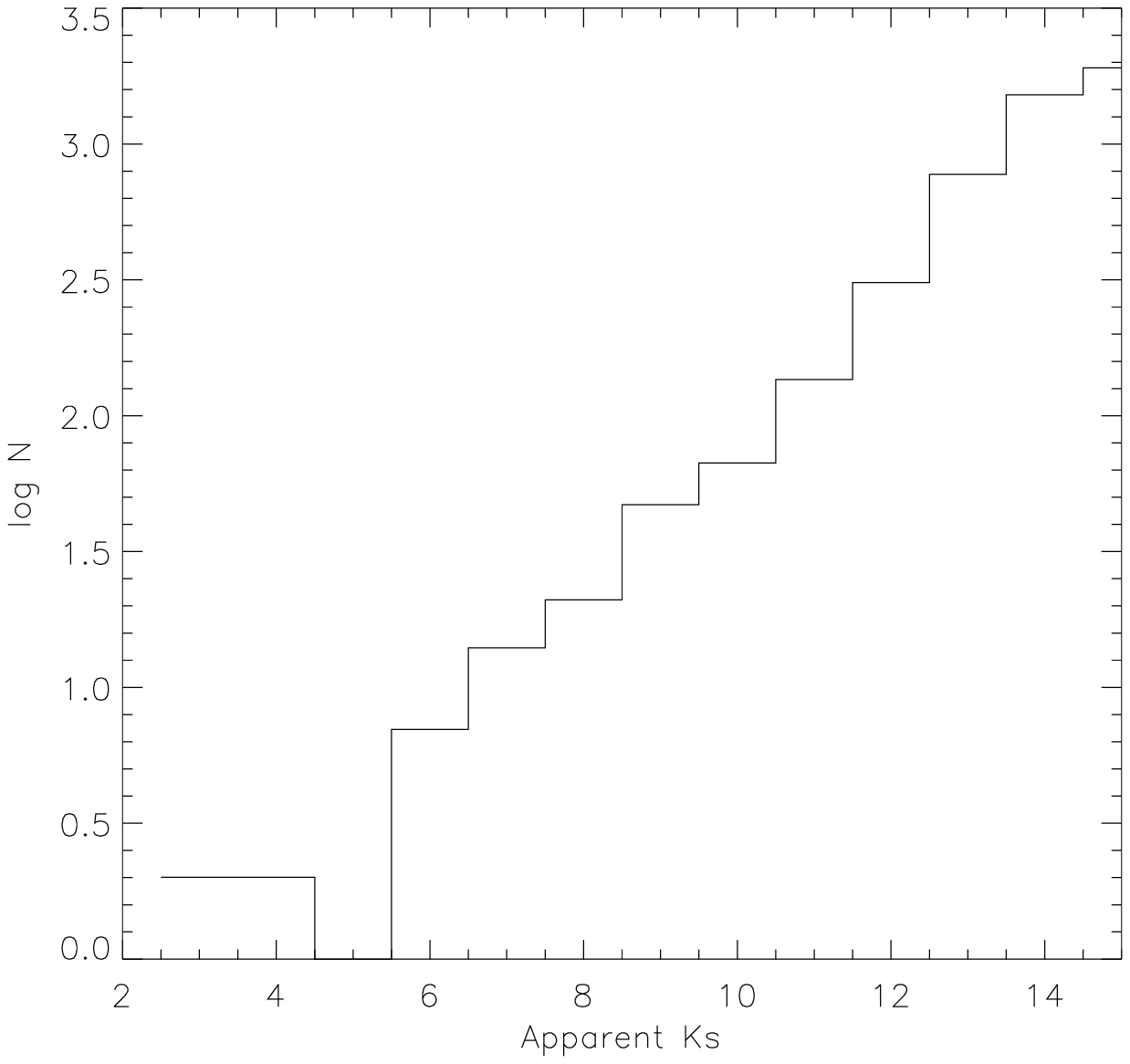]{The KLF of NGC 2244 after correction of the background
contamination. The KLF is found to match well a power law distribution and
gives a slope of 0.30.}

\newpage
\plotone{f1.ps}
\newpage
\plotone{f2a.ps}
\plotone{f2b.ps}
\newpage
\plotone{f3.ps}
\newpage
\plotone{f4a.ps}
\plotone{f4b.ps}
\newpage
\plotone{f5a.ps}
\plotone{f5b.ps}
\newpage
\plotone{f6a.ps}
\epsscale{0.6}
\plotone{f6b_color.ps}
\newpage
\epsscale{1.0}
\plotone{f7.ps}
\newpage
\plotone{f8.ps}
\plotone{f9.ps}

\end{document}